\documentclass[conference]{IEEEtran}

\usepackage{ifpdf}
\usepackage{cite}
\usepackage{color}

\ifCLASSINFOpdf
\else
\usepackage[dvips]{graphicx}
\fi

\usepackage[cmex10]{amsmath}
\usepackage{array}
\usepackage{mdwmath}
\usepackage{mdwtab}
\usepackage{amssymb}
\usepackage[font=footnotesize]{subfig}
\usepackage{fixltx2e}
\usepackage{url}

\begin{document}
\title{Surface-Roughness-Limited Mean Free Path\\ in Si Nanowire FETs}
\author{\IEEEauthorblockN{Hyo-Eun Jung and Mincheol Shin}\\
\IEEEauthorblockA{Dept. of Electrical Engineering,\\
Korea Advanced Institute of Science and Technology,
Daejeon 305-732, Rep. of Korea\\
Email: junghe@kaist.ac.kr, mshin@kaist.ac.kr}
 }

\maketitle

\begin{abstract}
The mean free path (MFP) in silicon nanowire field effect transistors limited by surface roughness scattering (SRS) is calculated with the non-perturbative approach utilizing the non-equilibrium Green's function method. The entrance scattering effect associated with finiteness of the channel length is identified and a method to eliminate it in the calculation of the MFP is developed. The behavior of the MFP with respect to  channel length (L), channel width (W), and the root-mean-square (RMS) of the surface roughness is investigated extensively. Our major findings are that the single parameter, RMS/W, can be used as a good measure for the strength of the SRS effects and that the overall characteristics of the MFP are determined by the parameter. In particular, the MFP exponentially decreases with the increase of RMS/W and the MFP versus the gate electric field shows a distinctively different behavior depending on whether the strength of the SRS effects measured by RMS/W is smaller or greater than 0.06.\\
\end{abstract}

\begin{IEEEkeywords}
nanowire, MOSFET, nanowire field effect transistor, quantum transport, mobility, mean free path, surface roughness, non-equilibrium Green's function.
\end{IEEEkeywords}
\section{INTRODUCTION}
The Si nanowire field effect transistors (SNWFETs) with all around gates have recently emerged as a promising device to replace the conventional planar metal oxide semiconductor field effect transistors (MOSFETs) as the latter face the physical and technological challenges in their scaling-down\cite{samsung,Singh,Jean}.
The SNWFETs have the merit of good electrostatic control and hence can deliver high on-state currents.
As their overall performance is determined by the aspect ratio of the channel length ($L$) and channel width ($W$)\cite{Mshin}, it may be desirable to utilize thinner and thinner nanowires.
However, rough interfaces between Si and dielectric are unavoidably generated during the fabrication processes, and currents become limited by surface roughness scattering (SRS)
as the nanowire cross-sectional size is reduced below $5$ nm\cite{Basu,Gamiz}.
\begin{figure}[t]
\vspace{0cm}
{\centering
\resizebox*{3.0in}{2.2in}{\includegraphics{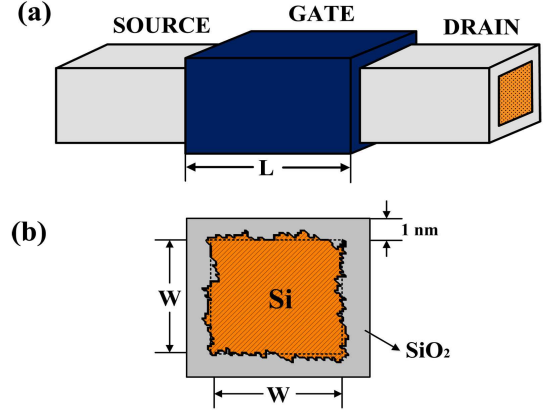}}
\par}
\vspace{-0.15cm} \caption{\footnotesize (a) Schematic diagram of the silicon nanowire field effect transistor with all-around gates. (b) Cross section view at the center of the channel with rough interfaces.}\label{fig:Fig1device}
\end{figure}

The SRS effects in SNWFETs have been a subject of intense theoretical and computational studies.
Most of the studies have been performed with the semiclassical approaches such as the Kubo-Greenwood approach\cite{Dura,SJin2007,Barraud},
the multi-subband Monte Carlo technique\cite{Asenov,Ramayya}, and the direct solution of the one-dimensional Boltzmann transport equation\cite{SJin2008,Lenzi}. The Schrodinger-Poisson equations are self-consistently solved in the approaches assuming an infinitely extended homogeneous nanowire, before the nanowire subband information is extracted and fed to the semiclassical procedures.
There have also been a few studies based the non-equilibrium Green's function (NEGF) method\cite{Asenov,Rogdakis,Wang,Poli,Buran,Martinez,SKim2011},
where coherent electron transport across the channel region with rough Si/dielectric interfaces is considered fully quantum-mechanically.
In the non-perturbative approach, the screening effect, electron phase coherence, and the mode mixing effect, among others, are naturally considered whereas these are hard to be implemented directly or disregarded entirely in the semiclassical approaches.

Most of the NEGF studies have so far focused on the low field electron mobility in nanowires. However, the mobility in the device with a finite channel length is not a proper measure for the device performance, because the mobility increases as the channel length increases whereas there should be more detrimental scattering. As Gnani has pointed out, the mean free path (MFP) should be a better indicator of the device performance\cite{Gnani}.

In this work, using the non-perturbative NEGF approach, we have for the first time investigated the SR-limited MFP in SNWFETs. Novelty and contributions of this work are:
$1$) we have devised a systematic way to calculate the MFP in the device with a finite channel length,
$2$) we have identified the entrance scattering effect which is unavoidable in the device with a source/channel/drain structure, estimated its strength, and eliminated it from the MFP calculation,
and $3$) we have thoroughly investigated the behavior of the MFP with a particular emphasis on its dependence on the strength of the SRS effects. These will be discussed in detail in the following sections.\\

\section{Methodology}

\subsection{Simulated devices and Hamiltonian}
The simulated devices in this work are three-dimensional ($3$D) rectangular silicon nanowire FETs with all-around gates as shown in Fig.\ \ref{fig:Fig1device}.
The source and drain regions are heavily $n$-doped with doping concentration of $10^{20}$cm$^{-3}$ whereas the channel is intrinsic.
The SiO$_{2}$ oxide thickness is assumed to be $1$ nm. To describe the electron transport in the conduction band, the effective mass Hamiltonian was used,
with the effective masses being scaled according to the tight-binding calculation\cite{Wang_IEEE}.
All the six equivalent valleys are included in the simulations.
\subsection{Surface roughness generation}
The surface roughness (SR) was realized at four Si/SiO$_{2}$ interfaces in the channel region only.
SR was generated according to the exponentially decaying autocorrelation function expressed as
\begin{equation}
C(\vec{r}) = \Delta_{m}^{2}\exp(-\sqrt{2}r/L_{m}),
\end{equation} \\
where $ \Delta_{m}$ is the root-mean square (RMS) fluctuation of the roughness,
$L_{m}$ is the correlation length, and $r = |\vec{r}|$ is the two-dimensional distance between two points on the interface.
By Fourier-transforming the autocorrelation function to obtain the power density spectrum (PDS), multiplying a random phase factor to the square root value of PDS,
and inverse Fourier-transforming the randomized PDS, instances of rough Si/SiO$_{2}$ interfaces were randomly generated\cite{Poli,Goodnick}.
\subsection{NEGF method}
We have employed the NEGF method to calculate the charge density and the current in the devices with rough surfaces in the channel region.
Self-consistent calculations coupling the charge density from the NEGF equations and the potential from Poisson's equation were carried out.
The methodology is well established in the literature\cite{Mshin_IEEE,SJin2008}.
\begin{figure}[t!]
\vspace{0cm}
 {\centering
\resizebox*{3.4in}{4.0in}{\includegraphics{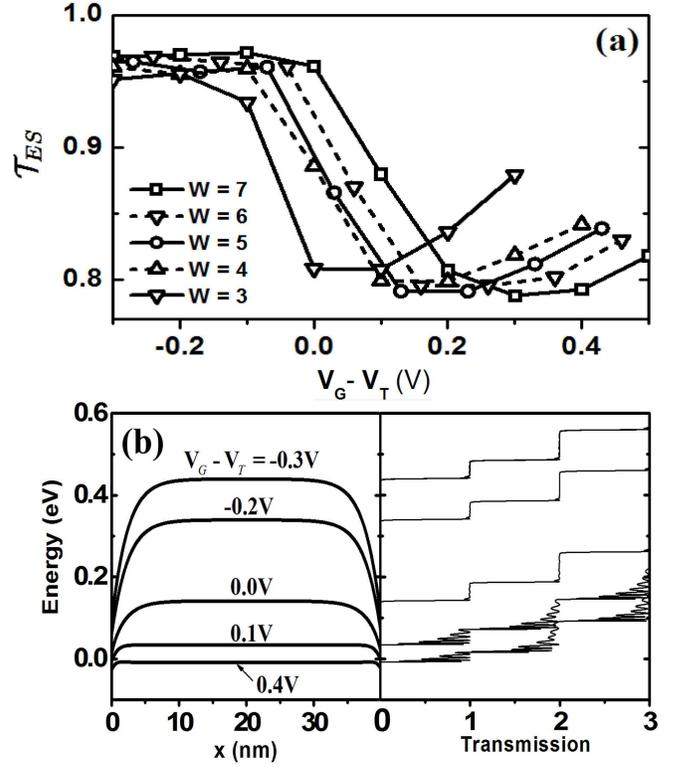}}
\par}
\vspace{0.0cm} \caption{\footnotesize (a) The effective transmission coefficient ${\cal{T}}_{ES}$ for different wire widths. (b) The conduction-band profiles (1st subband) for different gate voltages and their corresponding transmission functions. Device dimensions are $W$ = $5$ nm and $L$ = $40$ nm. }\label{fig:Fig2potAndte}
\end{figure}

The NEGF procedure with the rough surfaces is the same as the standard non-diffusive calculation procedure.
There is no need to calculate the less-than or greater-than Green's functions, because the surface roughness scattering (SRS) is not treated by
including the self-energy in the Green's function. Rather, the conduction band edge $E_{C}(\vec{r})$ at the Si/SiO$_{2}$ interfaces is spatially varied in
accordance with the generated rough surfaces.
In other words, in the Hamiltonian
\begin{equation}
H = H_{0} + E_{C}(\vec{r}) - q_{0}\phi(\vec{r}),
\end{equation}
where $H_{0}$ is the kinetic part of the Hamiltonian with the inverse effective mass tensor which is not altered by SR, $E_{C}(\vec{r})$ assumes the value of
of 0.56 eV or 3.63 eV, if $ \vec{r} $ corresponds to silicon or oxide points, respectively. In the Poisson's equation,
\begin{equation}
\overrightarrow{\nabla}\cdot(\epsilon(\vec{r})\overrightarrow{\nabla}\phi(\vec{r})) = q_{0}(n(\vec{r})-N_{D}) ,
\end{equation}
$\epsilon(\vec{r})$ is also spatially varying in the same manner as $E_{C}(\vec{r}) $.

In this work, the coupled mode space approach was used \cite{Mathieu,Wang_JAP} with the number of modes ranging from $30$ to $50$ depending on the cross-sectional width of nanowire.
Including higher modes little changed the outcome of the calculations, which validates that the used number of modes is sufficient. \\

\section{ENTRANCE SCATTERING EFFECT}
In the nanowire FET in Fig.\ \ref{fig:Fig1device}, electrons experience quantum mechanical scattering by the channel potential barrier as they enter the channel region.
The channel potential barrier is present due to the fact that the source/drain and channel regions are doped with different doping concentrations.
The entrance scattering effect (ESE) takes place even in the ballisticity condition (no rough surfaces in the channel region) and should be separated out
from the calculated mean free path. The method to isolate the pure contributions from the channel region is elaborated in section IV.

In the nanowire device under consideration, the ballistic current $I_{b}$ can be expressed as, at low $V_{DS}$,
\begin{equation}
I_{b} = \frac{2q_{0}^{2}V_{DS}}{h}\sum_{k}{\cal{T}}_{k}{\cal{F}}_{-1}(\frac{E_{F}-E_{k}}{k_{B}T}), \label{eq:Ib}
\end{equation}\\
where ${\cal{F}}_{m}(x)$ is the $m$-$th$ order modified Fermi integral,
$E_{k}$ is the energy level of subband $k$ in the channel region, $E_{F}$ is the Fermi energy, and ${\cal{T}}_{k}$ is the transmission probability in subband $k$ that accounts for
backscattering due to ESE.
If the source/drain regions are removed and the same cross-section of the channel region is repeated indefinitely, there is no ESE and
${\cal{T}}_{k} = 1$ in (\ref{eq:Ib}). Let us denote the ballistic current in the infinitely long homogeneous nanowire as $I_{b\infty}$.
We write
\begin{equation}
I_{b} = {\cal{T}}_{ES} I_{b\infty}, \label{eq:Ib2}
\end{equation}
where the effective transmission coefficient ${\cal{T}}_{ES}$ is introduced as a measure for the degree of ESE in the nanowire device.
\begin{figure}[t!]
\vspace{-0.6cm}
\hspace{0cm}
{ \centering
\resizebox*{3.8in}{2.9in}{\includegraphics{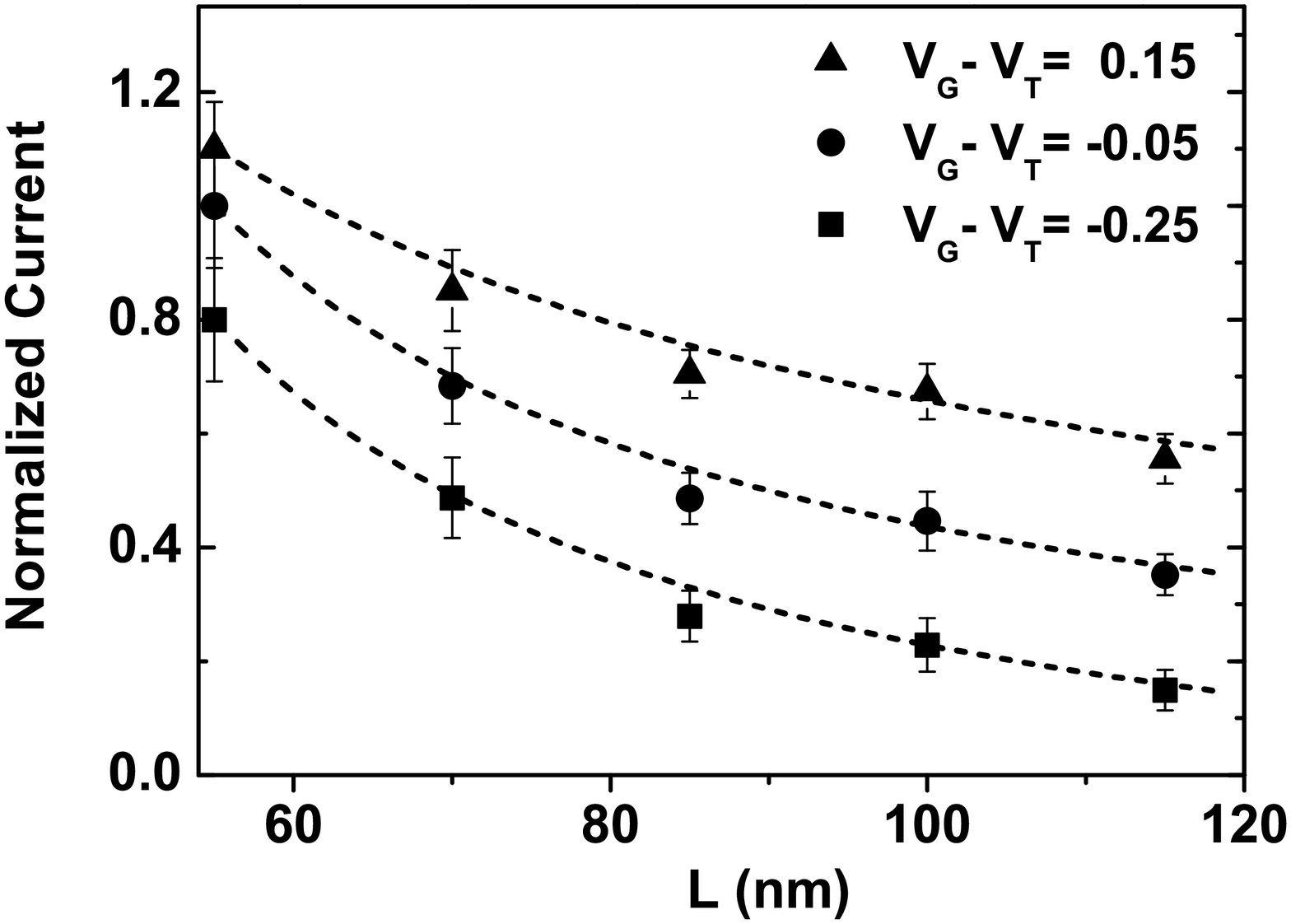}}
\par}
\vspace{-0.2cm} \caption{\footnotesize The drain current versus channel length $L$ for different gate voltages. The currents are normalized by the current of $L$ = 55 nm. Dashed lines are the fit to the function of the form $A/(B+L)$. The curves for $V_{G}-V_{T}$ = -0.25 V and 0.15 V are shifted by -0.2 and 0.1, respectively, for a clearer view. Device width and SR parameters are $W$ = $3$ nm, $\Delta_{m}$ = $0.3$ nm, and $L_{m}$ = $1.0$ nm. }\label{fig:norm_I}
\end{figure}
\begin{figure}[h!]
\vspace{-0.6cm}
\hspace{-0.02cm}
{ \centering
\resizebox*{3.95in}{2.9in}{\includegraphics{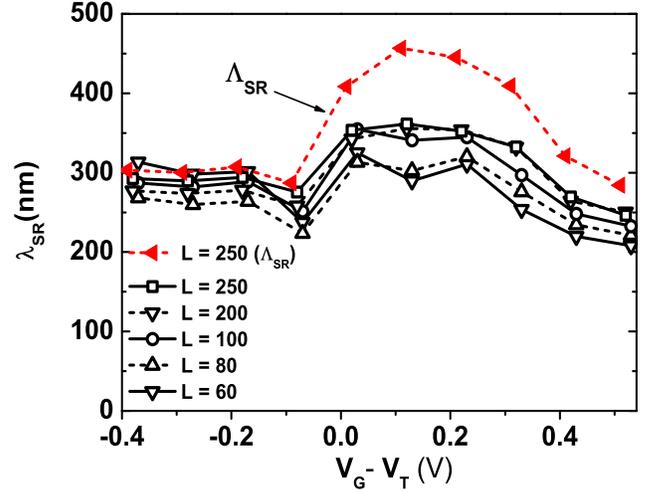}}
\par}
\vspace{-0.2cm} \caption{\footnotesize The mean free path $\lambda_{SR}$ for different channel lengths. $\Lambda_{SR}$ as defined in (\ref{eq:BigLambda}) is also displayed.  Device width and SR parameters are $W$ = $5$ nm, $\Delta_{m}$ = $0.2$ nm, and $L_{m}$ = $1.0$ nm. }\label{fig:mfpL}
\end{figure}

Fig.\ \ref{fig:Fig2potAndte} (a) shows ${\cal{T}}_{ES}$ versus $V_{G}$ for various $W$'s, where it is shown that ${\cal{T}}_{ES}$ is as low as about $0.8$ and its dependence on
$V_{G}$ is not sensitive to $W$.
As shown in Fig.\ \ref{fig:Fig2potAndte} (b), the channel potential barrier is so high in the subthreshold regime that an electron is either completely reflected or completely transmitted depending on whether its energy is below or above the potential barrier, except for a very narrow window of energy located just above the potential barrier where it experiences the quantum-mechanical backscattering.
So ${\cal{T}}_{ES}$ is close to 1 in the subthreshold regime. Around $V_{T}$, the potential barrier is lowered and at the same time its shape near the ends of the channel becomes sharper so that the quantum mechanical backscattering starts to occur seriously.
In other words, the above-mentioned energy window broadens, resulting in a sharp decrease of ${\cal{T}}_{ES}$.
If the gate voltage is further increased, ${\cal{T}}_{ES}$ reaches a minima and then slightly increases, because the channel potential barrier is sufficiently lowered to result in lesser backscattering.
\\
\section{MEAN FREE PATH}
In devices with finite channel length, the effective mobility is no longer a good indicator of device performance,
as it tends to zero as the ballistical condition is approached whereas the ON current increases \cite{Gnani}.
Instead, the mean free path (MFP) that is directly related to electron transmission in the device is better suited as the indicator of device performance.
In this work, the SR-limited MFP is calculated as follows.

At a low drain bias, we relate $I$ with $I_{b\infty}$ via
\begin{equation}
I = {\cal{T}} I_{b\infty}, \label{eq:IT}
\end{equation}
where ${\cal{T}}$ is the overall transmission coefficient in the device. ${\cal{T}}$ consists of contributions from the entrance scattering (${\cal{T}}_{ES})$ and the
surface roughness scattering (${\cal{T}}_{SR}$).
In an attempt to isolate ${\cal{T}}_{SR}$, we write
\begin{figure}[t!]
\vspace{-0.7cm}
 {\centering
\resizebox*{3.9in}{2.85in}{\includegraphics{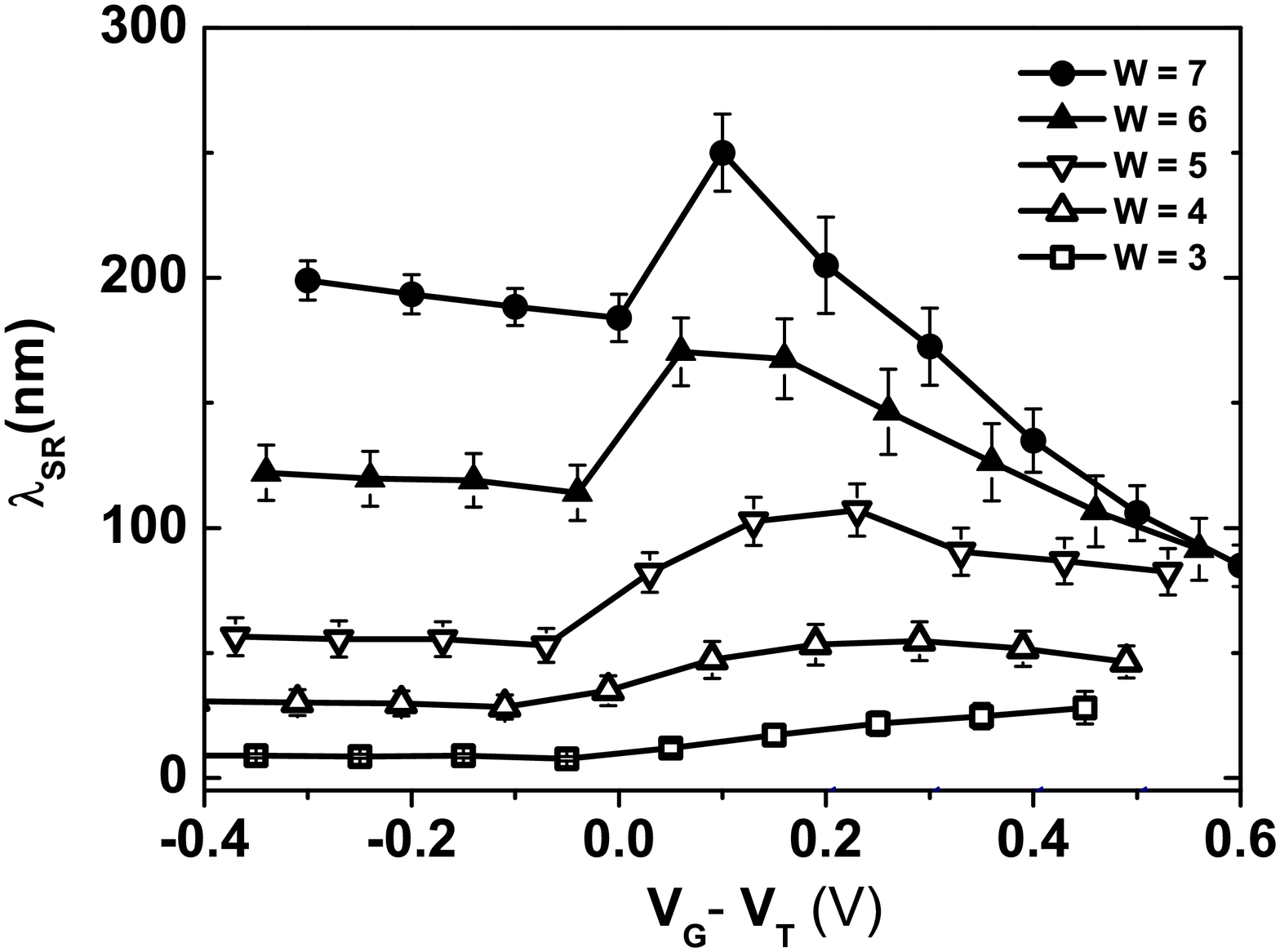}}
\par}
\vspace{-0.2cm} \caption{\footnotesize The mean free path $\lambda_{SR}$ for different wire widths. The channel lengths were fixed at $L$ = $100$ nm for all $W$'s. SR parameters are $\Delta_{m} = 0.3$ nm and $L_{m} = 1.0$ nm. Mean value and sample standard deviation (dispersion bars) are reported.}\label{fig:mfpW}
\end{figure}
\begin{equation}
\frac{1-{\cal{T}}}{{\cal{T}}} = \frac{1-{\cal{T}}_{ES}}{{\cal{T}}_{ES}} + \frac{1-{\cal{T}}_{SR}}{{\cal{T}}_{SR}}, \label{eq:1/T}
\end{equation}\\
which is written based on the simple picture that the two sources of scattering are connected serially\cite{Datta}. Using the defining relationship for the SR-limited MFP $\lambda_{SR}$,
\begin{equation}
{\cal{T}}_{SR} = \frac{\lambda_{SR}}{\lambda_{SR}+L}, \label{eq:T_SR}
\end{equation}
we finally obtain
\begin{equation}
\lambda_{SR} = \frac{L}{{\cal{T}}^{-1}-{\cal{T}}_{ES}^{-1}}.
\end{equation}
Notice that in defining ${\cal{T}}$ in (\ref{eq:IT}), $I_{b\infty}$ is used instead of $I_{b}$.
If the relationship $I = {\cal{T}}^{\prime}I_{b}$ is used instead of (\ref{eq:IT}), it implies that ESE is disregarded, and we may write
\begin{equation}
{\cal{T}}^{\prime} = \frac{\Lambda_{SR}}{\Lambda_{SR}+L}.  \label{eq:T_prm}
\end{equation}
The MFP $\Lambda_{SR}$ of the above equation is related to $\lambda_{SR}$ as:
\begin{equation}
\Lambda_{SR} = \frac{1}{{\cal{T}}_{ES}}\lambda_{SR}. \label{eq:BigLambda}
\end{equation}
That is, the MFP is multiplied by the factor $1/{\cal{T}}_{ES}$, which leads to an unphysical peak as demonstrated below.

Before proceeding further, we first show $I$ as a function of $L$ in Fig.\ \ref{fig:norm_I}, which verifies that diffusive transport expressed by Eqs.\ (\ref{eq:T_SR}) or (\ref{eq:T_prm}) actually takes place due to SR. Note that we have deliberately chosen the device width and the SR parameters in the figure so that resultant MFP is as small as about 10 nm (consult Fig.\ \ref{fig:mfpR}(a)).

Fig.\ \ref{fig:mfpL} shows thus-calculated $\lambda_{SR}$ for SNWFETs of $W= 5$ nm. For $L$ varying from $60$ nm to $250$ nm, $\lambda_{SR}$ seems to depend on $L$ weakly.
The MFP calculated by using (\ref{eq:BigLambda}) is also displayed in the figure: $\Lambda_{SR}$ shows a pronounced peak around $V_{T}$ when there is no physical reason for the feature.
This supports that $\lambda_{SR}$ is a more natural choice for the MFP than $\Lambda_{SR}$.
In the $\lambda_{SR}$ versus $V_{G}$ curve of Fig.\ \ref{fig:mfpL}, however, there is still some hint of a peak around $V_{T}$.
We believe that this is an artifact which is originated from the imperfect nature in the process of isolating the pure SR contribution in the channel region ((\ref{eq:1/T})).
As a matter of fact, (\ref{eq:1/T}) should hold better for greater $L$ because the assumption that the two sources of scattering (ES and SR) are incoherently connected becomes more valid.
\begin{figure}[t!]
\vspace{0.1cm}
 {\centering
\resizebox*{3.0in}{3.4in}{\includegraphics{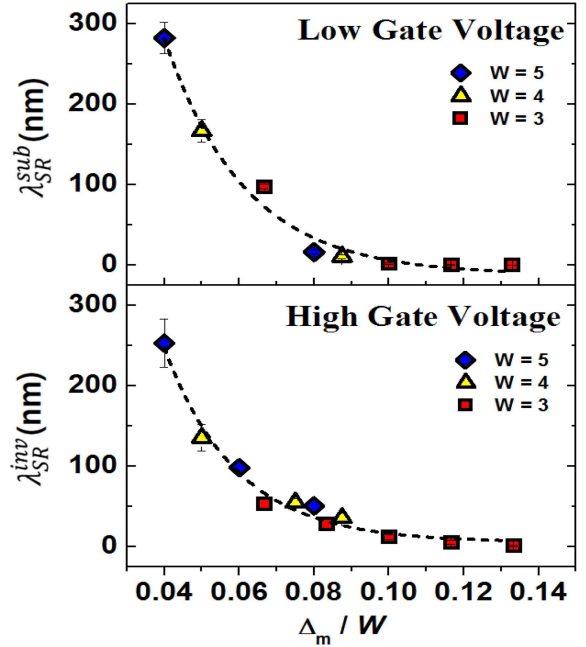}}
\par}
\vspace{-0.2cm} \caption{\footnotesize (Top) The mean free path $\lambda_{SR}^{sub}$ and (Bottom) $\lambda_{SR}^{inv}$ evaluated at $V_{G}-V_{T} = -0.3$ V and $0.4$ V versus $\Delta_{m}/W$, respectively. Both $W$ and $\Delta_{m}$ were varied and $L_{m}$ = 1 nm. The data from $W$ = $3$, $4$, and $5$ nm are represented by squares, triangles, and diamonds, respectively. The channel lengths were fixed to $100$ nm for all calculations. The dashed lines represent data fit to an exponentially decaying function. }\label{fig:mfpEX}
\end{figure}

Fig.\ \ref{fig:mfpW} shows the MFP for different $W$'s when $L$ is fixed at $100$ nm.
The MFP decreases as $W$ becomes smaller, as expected, but the above-mentioned peaky feature near $V_{T}$ is more pronounced at larger $W$.
Hereafter let us set aside the unphysical peaky feature near $V_{T}$ and concentrate on the MFP in the subthreshold region ($\lambda_{SR}^{sub}$) and the inversion region ($\lambda_{SR}^{inv}$).

Our key findings in this work with regard to the MFP's are: $1$) the dimensionless parameter $\Delta_{m}/W$ can be used as a good measure of the SRS strength, $2$) the MFP's exponentially decrease with $\Delta_{m}/W$, and $3$) $\lambda_{SR}^{inv}$ becomes greater than $\lambda_{SR}^{sub}$ if $\Delta_{m}/W\gtrsim 0.06$.

Firstly, both $\lambda_{SR}^{sub}$ and $\lambda_{SR}^{inv}$ exponentially decrease with $\Delta_{m}/W$ as shown in Fig.\ \ref{fig:mfpEX}, where $\lambda_{SR}$'s were calculated
for all the possible combinations of $W$ = $3$, $4$, and $5$ nm and $\Delta_{m}$ is $0.1\sim0.4$ nm.
$\lambda_{SR}^{sub}$ and $\lambda_{SR}^{inv}$ were evaluated at $V_{G}-V_{T}$ = $-0.3$ V and $0.4$ V, respectively.
Notice that $W$ is restricted to below $5$ nm here. In particular, $\lambda_{SR}^{inv}$ for $W \geq 5$ nm converges to the almost the same value, as shown in Fig.\ \ref{fig:mfpW}.
\begin{figure}[t!]
\vspace{-0.4cm}
 {\centering
\resizebox*{3.0in}{3.45in}{\includegraphics{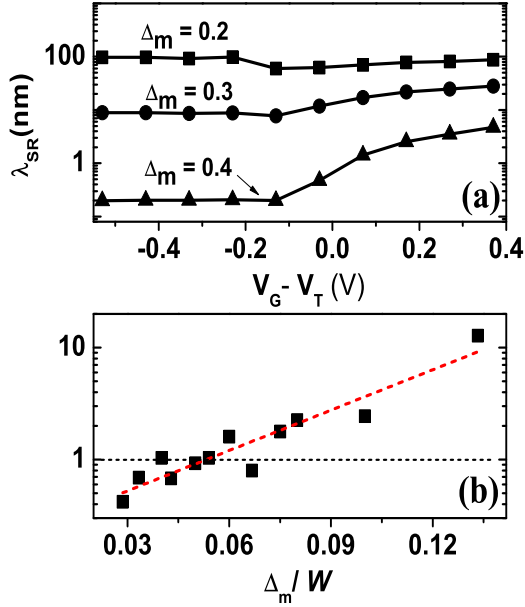}}
\par}
\vspace{-0.3cm} \caption{\footnotesize (a) The mean free path $\lambda_{SR}$ versus $V_{G}$ for $W = 3$ nm and $L = 100$ nm, as $\Delta_{m}$ is changed from $0.2$ to $0.4$ nm and $L_{m}$ = 1 nm.
(b) The ratio $\lambda_{SR}^{inv}/\lambda_{SR}^{sub}$ versus $\Delta_{m}/W$. }\label{fig:mfpR}
\end{figure}

The fact that the MFP can be described by a single parameter $\Delta_{m}/W$, even with different $W$'s, is quite interesting.
For the size quantization effects associated with channel width, such as the degree of wave-function confinement, the number of subbands,
and the degree of the mixing between subbands, are expected to result in the MFP's dependence on $W$ in a complicate way.
Nevertheless, our calculations show the MFP's exponential dependence on $\Delta_{m}/W$.
This should give a useful guide for the estimation of $\lambda_{SR}$ in Si nanowire devices.

Secondly, we have found that, if the SRS effects become sufficiently strong, $\lambda_{SR}$ increases with the increase of $V_{G}$ and,
as the consequence, $\lambda_{SR}^{inv} > \lambda_{SR}^{sub}$.
This is in sharp contrast to the usual case where  $\lambda_{SR}$ should decrease with the increase of $V_{G}$, because more intense scattering is expected to occur at higher gate voltages.
In our calculations, the latter is observed for the cases of the SRS effects being relatively weak. See $\lambda_{SR}$ for $W$ = $5$ nm ($\Delta_{m}$ = $0.2$ nm) in Fig.\ \ref{fig:mfpL}
and that for $W$ = $6$ and $7$ nm ($\Delta_{m}$ = $0.3$ nm) in Fig.\ \ref{fig:mfpW}, respectively.
However, as $W$ becomes smaller in Fig.\ \ref{fig:mfpW}, the usual behavior is not seen any longer and $\lambda_{SR}^{inv}$ becomes greater than $\lambda_{SR}^{sub}$.

We argue that the ratio $\lambda_{SR}^{inv} / \lambda_{SR}^{sub}$ has to do with the strength of the SRS effects and that $\Delta_{m}/W$ is a good measure for the strength.
Obviously, the SRS effects should become stronger as $W$ becomes smaller for fixed $\Delta_{m}$(case $A$) or as $\Delta_{m}$ becomes bigger for fixed $W$(case $B$).
For case $A$,  $\Delta_{m}$ is fixed to $0.3$ nm and $W$ is varied as shown in Fig.\ \ref{fig:mfpW}, and we observe that $\lambda_{SR}^{inv} / \lambda_{SR}^{sub}$ becomes
bigger than $1$ if $\Delta_{m}/W \geq 0.06$. For case $B$, $W$ is fixed to $3$ nm and $\Delta_{m}$ is varied as shown in Fig.\ \ref{fig:mfpR} (a).
For $\Delta_{m}$ = $0.2$ nm in the figure,  $\lambda_{SR}$ already starts to ascend in the inversion regime, although quite slowly.
As $\Delta_{m}$ is increased to $0.3$ nm and then to $0.4$ nm, $\lambda_{SR}$ ascends more and more steeply and $\lambda_{SR}^{inv}$ becomes as big as about ten times of $\lambda_{SR}^{sub}$ for $\Delta_{m}$ = $0.4$ nm. (Also compare the curves of $W$ = $5$ nm in Fig.\ \ref{fig:mfpL} and Fig.\ \ref{fig:mfpW}, where $\Delta_{m}$ is changed from $0.2$ nm to $0.3$ nm.)

If we combine cases $A$ and $B$ and plot $\lambda_{SR}^{inv} / \lambda_{SR}^{sub}$ versus $\Delta_{m}/W$ in Fig.\ \ref{fig:mfpR} (b), we observe that the ratio $\lambda_{SR}^{inv} / \lambda_{SR}^{sub}$ generally increases with the increase of $\Delta_{m}/W$, although there are small fluctuations. $\lambda_{SR}^{inv} / \lambda_{SR}^{sub}$ crosses the line of $1$ at $\Delta_{m}/W \thicksim 0.06$ in the figure. Thus we conclude that the characteristics of the SRS effects in Si nanowires change at $\Delta_{m}/W \thicksim 0.06$.
\begin{figure}[t!]
\vspace{-0.5cm}
 {\centering
\resizebox*{3.2in}{3.5in}{\includegraphics{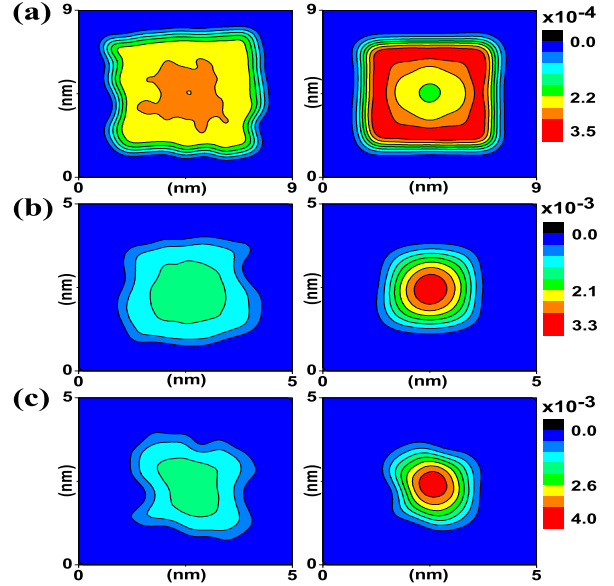}}
\par}
\vspace{-0.5cm} \caption{\footnotesize The cross-sectional electron density profiles for low $V_{G}$ (left column) and high $V_{G}$ (right column), for (a) $W$ = $7$ nm, $\Delta_{m} = 0.2$ nm and (b) $W$ = $3$ nm, $\Delta_{m} = 0.2$ nm, and (c) $W$ = $3$ nm, $\Delta_{m} = 0.4$ nm. Electron densities are normalized by the total charge density of the cross-section. }\label{fig:3DNd}
\end{figure}

That $\lambda_{SR}^{inv}$ becomes greater than $\lambda_{SR}^{sub}$ for the strong SRS effects can be qualitatively explained as follows. In Figs.\ \ref{fig:3DNd} (b) and (c), the cross-sectional charge densities for the cases of $\Delta_{m}$ = $0.2$ and $0.4$ nm are compared at low and high gate voltages ($W$ = $3$ nm). In the charge density profiles at low gate voltage, the Si/SiO$_{2}$ interfaces that fluctuate in proportion to $\Delta_{m}$ can be directly observed. In the charge density profiles at high gate voltage, however, the degree of the interface fluctuations seems to be almost invariant to the increase of the surface roughness. The latter is due to the wave function confinement effect which is of particular importance for nanowires of $W \leq 5$ nm, and consequently, the electrons are concentrated to the center of the nanowire as seen in the figure \cite{Mshin,Poli} . Hence the `effective' surface roughness that felt by the electrons is less at high gate voltage than at low gate voltage. This effect intensifies as $\Delta_{m}$ is increased or $W$ is decreased, resulting in the relationship between $\lambda_{SR}^{inv} / \lambda_{SR}^{sub}$ and $\Delta_{m}/W$ shown in Fig.\ \ref{fig:mfpR} (b). On the other hand, for nanowires of $W > 5$ nm, the confinement effect is weak so the electrons are pushed to the nanowire surfaces at high gate voltage, resulting in the expected behavior that the MFP decreases with the increase of the gate voltage as shown in Fig.\ \ref{fig:3DNd} (a).\\

\section{Conclusion}
In this work, we have systematically devised the methodology for the calculation of the mean free path in the nanowire field effect transistors with finite channel length.
The parasitic entrance scattering effects were identified and carefully eliminated from the MFP calculations.
Non-perturbative approach based on the non-equilibrium Green's function method was used to address the SRS effects in the nanowire transistors.
The methodology developed in this work can be applied to any scattering mechanism, although only SRS is considered here.

We have found that the SR-limited MFP can be well described by a function which depends on a single parameter $\Delta_{m}/W$.
In particular, the MFP in the subthreshold regime exponentially decreases with the increase of $\Delta_{m}/W$. For nanowires of $W \leq 5$ nm,
the MFP in the inversion regime also exponentially decreases with $\Delta_{m}/W$.

We have also found that the dimensionless parameter $\Delta_{m}/W$ can be used as a good measure of the SRS strength. If $\Delta_{m}/W < 0.06$,
one may expect that the SRS effects should be relatively weak, resulting in the usual behavior that the MFP decreases with the increase of the gate electric field.
If $\Delta_{m}/W > 0.06$, on the other hand, the SRS effects can be regarded to be strong, resulting in the extraordinary behavior that the MFP increases with the increase
of the gate electric field. Our results should give a useful guide for the estimation of MFP and the strength of the SRS effects in the nanowire devices.\\

\section*{Acknowledgment}
This research was supported by the Pioneer Research Center Program and the Basic Science Research Program thorough the National Research Foundation of Korea (NRF) Funded by the Ministry of Education, Science and Technology (Grant No. 2012-0000459 and No. 2012-0002120)\\

\end{document}